\documentclass[12pt,preprint]{aastex}

\newcommand {\marti}[1]{{#1}${\mbox{\scriptsize{a}}\over\ }$}

\shorttitle{Jets Produced by Intense Laser}
\shortauthors{Mizuta et al.}

\begin{document}
\title{Numerical Analysis of Jets Produced by Intense Laser}

\author{Akira Mizuta, Shoichi Yamada, and Hideaki Takabe}

\affil{Institute of Laser Engineering, Osaka University,
Suita, Osaka, 565-0871, Japan and
Graduate School of Science, Osaka University,
Toyonaka, Osaka, 565-0043, Japan}

\begin{abstract}
In this paper we present a numerical study of plasma
jets produced by intense laser matter interactions.
Through this study we hope to
better understand astrophysical jets
and their recent experimental simulations in the laboratory.
We paid special attention to radiation cooling and
the interaction of the jet with ambient gas.
Four cases are presented in this paper;
two of them deal with the propagation of jets in vacuum,
while in the other two the propagation takes place in the ambient gas.
Available experimental results are
reproduced to good accuracy in the vacuum case.
For jets in ambient gas,
we find that the existence of the surrounding gas
confines the jet into a narrow cylindrical shape so that
both the density and temperature of the jet remain
high enough for effective radiation cooling.
As a result, a collimated plasma jet is formed
in these cases.
The dimensionless parameters characterizing the laboratory jets
and protostellar jets have overlapping domains.
We also discuss the cooling lengths for our model
and compare them with the corresponding values in the astrophysical jets.
A plasma jet in the ambient gas experiment
is proposed which is within the reach of present day technology,
and can be relevant to astrophysical phenomena.
\end{abstract}


\section{Introduction}
Astrophysical jets found in supersonic bipolar flows from black holes,
active galactic nuclei(AGN), protostars and similar phenomena 
are subject to active research.
Several problems concerning jets are remaining to be solved,
for example, why do the jets propagate such a long distance so stably.
Many numerical simulations of astrophysical jets
have been performed so far.
2D non-relativistic calculations have been done for AGN jets,
by \citet{Nor82}.
They found some dimensionless parameters relevant to
the propagation of jets over a long distance.
2D and 3D relativistic calculations have been carried out
by \citet{Marti97,Aloy99}.
Protostellar jets got more attention,
detailed numerical simulations have been performed by several
authors (see \citet{Reipurth97a,Reipurth97b,Reipurth97c,Bally01}
for observational data).
Since the radiation cooling and/or external magnetic field
are supposed to have very important effect
on the plasma jets propagation,
radiation hydrodynamic/magnetohydrodynamic codes are essential.
For example, \citet{Blondin90,Stone93} showed
with their radiation hydrodynamic codes
that the radiation cooling plays an important role for the collimation,
leading to the formation of a dense cold shell at the head of the jet.
\citet{Todo93} studied, by means of with MHD code the propagation of
jets from young stellar objects into ambient matter,
and showed that the azimuthal magnetic field of $\sim$
70 $\mu$G is strong enough to drive a helical kink instability.
Since then, many efforts have been put to simulations
in 2D and 3D with both radiation cooling and magnetic fields included
\citep{Cerqueira97,Frank98,Gardiner00,Frank00}.
The parametric researches that vary the form of the injection,
steady or pulsed jets, and the configuration of magnetic field
(helical or longitudal)
helped to clarify the systematics of jets.
In particular, it was shown that a helical magnetic field
can affect the morphology of the head of jets.
It was also demonstrated that 
Kelvin-Helmholtz instability is crucial for
the morphology of jets, and
linear as well as nonlinear stability analyses were carried out
with or without magnetic field
to characterize the influence of these factors on the jet shape
\citep{Xu00,Hardee97,StoneHar00,Ryu00}.

So far these studies are purely theoretical.
Recently, however, with the advance of femtosecond laser technology,
an experimental investigation of these astrophysical phenomena
in the laboratory seems to be possible \citep{Rem99,Rem00b}.
This new scientific field is called laboratory astrophysics
or laser astrophysics.
This progress enable the investigation of radiation hydrodynamics,
hydrodynamic instabilities, and similar astrophysical phenomena
in model experiments with intense lasers.
Such experiments might be able to give us a new and improved
insight into those phenomena which could not be obtained by
studies of extraterrestrial
observations or numerical simulations.
Some experiments along these lines
have already been carried out, in which
hydrodynamic instabilities and
the interactions of a strong shock with matter
have been measured.
Other experiments have been proposed \citep{Rem00a}.
Backup by numerical simulations
are indispensable, because they can connect between the processes
occurring in laser produced plasmas to
astrophysical phenomena that are the aim of our studies.
This paper is one of those attempts along this line.
In this paper, we will propose a different experiment
to understand astrophysical jets. 

\begin{figure}
\rotatebox{270}{\resizebox{6.0cm}{!}{\plotone{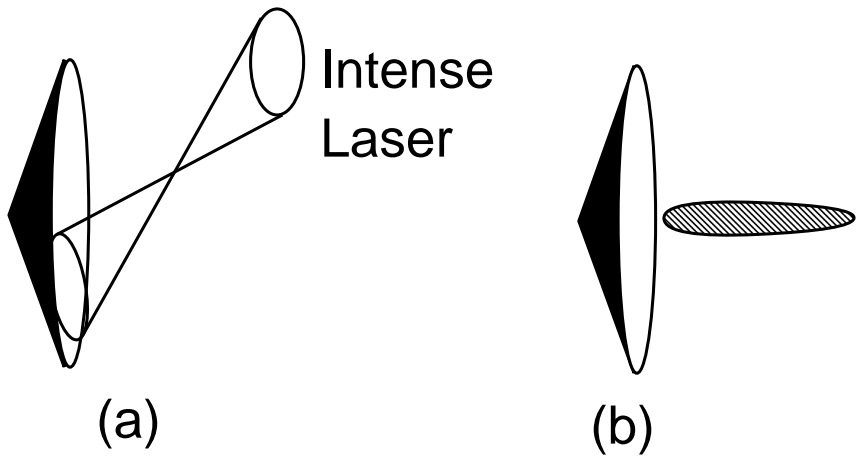}}}
\caption{'Cone' target. (a):Some intense lasers
 irradiate on the 'cone' target surface. (b):Ablation plasma will collimate
 and become a jet-like structure.\label{target}}
\end{figure}

Some experiments relevant to astrophysical jets have been performed
with intense lasers \citep{Far99,Shige00,Stone00}.
The so-called 'cone' target (Fig.\ref{target})
is employed in those experiments.
Another type of experiment has been recently done
by \citet{Lebedev01},
using a conical array of fine metric wires.
In general, intense laser matter interaction generates
high temperature and high density plasma
which starts to ablate immediately.
If the intensity of laser is $\sim 10^{14} \mbox{W cm}^{-2}$,
the expansion velocity of the plasma
reaches a few hundreds of kilometers per second.
This is of the same order as the velocity of protostellar jets.
This indicates that jets which realistically
mimic astrophysical ones can be produced in laboratory.
In these experiments, however, the plasma expands into vacuum,
and one can not the investigate influence of ambient gas,
which does exist for the astrophysical jets.
As a matter of fact, the ratio of the mass density
of the jet to that of the interstellar gas 
($\rho _{jet}/\rho_{ambient}$) is $1\sim 10$ for protostellar jets.

In this paper,
we investigate the propagation of
laser-produced jets not only in the vacuum but also
in the ambient gas using the 2D hydrodynamics code developed
recently \citep{mizuta01}.
Although the code has two versions,
relativistic and non-relativistic ones,
in the present paper we use only the non-relativistic code.

This paper is organized as follows.
In sections 2 and 3,
the basic formulas, the numerical method and the initial conditions
are presented.
The main results are shown in section 4.
We discuss some implications to
astrophysical jets in section 5
and give a short summary in section 6.

\section{Basic Equations and Numerical Method}
Assuming a plasma having axial symmetry 2D
Euler equation for perfect fluid is solved:
\begin{eqnarray}
\label{hozon}
{\partial \mbox{\boldmath $u$} \over \partial t}+
{1\over r}{\partial (r\mbox{\boldmath $f$}) \over \partial r}+
{\partial \mbox{\boldmath $g$}\over \partial z}=\mbox{\boldmath $s$},
\end{eqnarray}
where the conserved variable vector \mbox{\boldmath $u$}, 
the flux vectors \mbox{\boldmath $f$} and \mbox{\boldmath $g$}, and 
the source vector \mbox{\boldmath $s$} are defined as,
\begin{eqnarray}
\mbox{\boldmath $u$}=(\rho,\rho v_{r},\rho v_{z},E)^{T},\\
\mbox{\boldmath $f$}=
(\rho v_{r},\rho v_{r}^2+p,\rho v_{r}v_{z},(E+p)v_{r})^{T},\\
\mbox{\boldmath $g$}=
(\rho v_{z},\rho v_{r}v_{z},\rho v_{z}^2+p,(E+p)v_{z})^{T},\\
\mbox{\boldmath $s$}=(0,p/r,0,0)^{T},\\
E=\rho (\epsilon + (v_{r}^2+v_{z}^2)/2).
\end{eqnarray}
In the above equations,
 $\rho , v_{i}, p,$ and $\epsilon\ $are mass density, 
$i$-component of velocity, 
pressure, and specific internal energy, respectively.
We ignore viscosity and heat conduction of the fluid.
These assumptions are justified as follows.
As for the collisions,
the parameter $\delta=\lambda _{mfp}/L_{sys}$ defined as
the ratio of the collisional mean free path
$\lambda _{mfp}$ of electrons or ions
and, the characteristic length of the system $L_{sys}$,
is $10^{-4}\sim 10^{-6}$ for our models described below.
Hence a hydrodynamic description of the jets is appropriate.
Other dimensionless parameters characterizing
the dissipative processes are
the Reynolds number $Re \equiv r_{j}v_{j}/\nu$
which is a measure of the local viscosity
and the Peclet number $Pe \equiv r_{j}v_{j}/\nu_{h}$
which describes the heat conduction inside the jet.
Here $r_{j}$, $v_{j}$, $\nu$ and $\nu_{h}$ are
the radius and velocity of the jet,
the viscosity and thermal diffusivity coefficients respectively.
Criteria for the influence of these dissipative phenomena
on the flow are $Re \lesssim 1$ and $Pe \lesssim 1$.
In our models
we find $Re \gtrsim 10^9$ and $Pe \gtrsim 10^5$.
Thus the neglection of these dissipative processes is
well justified.
These numbers are evaluated at $t=2.0$ ns
at the center of the inflow region, see below.

An ideal gas equation is used in the computations,
\begin{eqnarray}
\label {eos}
p=\rho \epsilon (\gamma -1),
\end{eqnarray}
where $\gamma$ is the adiabatic exponent (=5/3 in this paper). 

We adopt the Marquina's flux formula \citep{Donat96} 
which is based on an approximate Riemann solver derived 
from the spectral decomposition of the Jacobian matrix of 
Euler equations.
We have developed a computational code
which solves Eq. (\ref{hozon}) and (\ref{eos})
The code uses
cylindrical coordinates with $300(r)\times 1500(z)$ grid points.
The accuracy of the code is the
second order in space due to a MUSCL method\citep{van77,van79}
and is the first order in time.

It is assumed that the plasma is optically thin,
so the radiation cooling effect is important.
In the cooling term $|J_{rad}|$, only for the bremsstrahlung
is included \citep{Zel66},
\begin{eqnarray}
\label{cooling}
|J_{rad}|=1.42\cdot 10^{-27}{Z^{*}}^2T^{1/2}n_{e}n_{ion}
{\mbox{erg cm}^{-3} {\mbox{s}}^{-1}},
\end{eqnarray}
where $Z^{*}$,$T$,$n_{e}$ and $n_{ion}$ are 
the average ionic charge, temperature and
number densities of electron and ion, respectively.
The temperature and average ionic charge are related to
the specific internal energy $\epsilon$ by,
\begin{eqnarray}
\label{ezt}
\epsilon ={3\over 2}\cdot {(Z^{*}+1)T\over m_{i}},
\end{eqnarray}
where $m_{i}$ is the mass of ion.
The average ionic charge state $Z^{*}$ and temperature $T$
were calculated iteratively from
local instantaneous density and specific internal energy,
as obtained from the hydrodynamic
calculation by using Eq.(\ref{ezt}) and approximate expression based on
the Thomas Fermi model for $Z^{*}$ \citep{sal98}.
The formulas for the line emission processes
were computed in the usual manner and
are not included here for simplicity.
As already mentioned the radiation transport
is also neglected in this paper.
Since we are interested mainly
in the qualitative effects of radiation cooling,
these approximations are adequate.

We have to distinguish the target matter from the ambient matter
in order to make the radiation cooling effective only for the former.
For this purpose we introduced another continuity
equation for the target matter:
\begin{eqnarray}
{\partial (\rho f) \over \partial t}+
{1\over r}{\partial (r\rho v_{r}f)\over \partial r}+
{\partial (\rho v_{z}f)\over \partial z}=0,
\end{eqnarray}
where $f$ is the fraction function.
This equation is solved simultaneously
with the hydrodynamic equations.
With $Z^{*}, T,$ and $f$ thus obtained,
the cooling term is calculated in a separate step.
For numerical convenience,
the radiation cooling term is turned off for very low temperatures,
less than 70 eV.
The radiation cooling term is neglected also in the vicinity
of the target, where the radiative processes are very complicated and
the basic picture of our model is not valid anyway.

\section{Initial Condition}
In order to simulate the experiments,
we put a 'cone' target at an end of the computational region.
At $t=0$, the laser deposits all its energy on the surface.
The depth of the surface is $\sim 10 \mu \mbox{m}$.
A Gaussian shaped laser pulse of duration of $100 \mbox{ps}$
was assumed.
This is much shorter than
the dynamical time scale, which is more than a few nanoseconds.
When the target made of gold is irradiated by intense laser of
$\sim 10^{14} {\mbox{W cm}^{-2}}$,
the temperature of the target
rises to  a few hundred eV up to 1 keV,
and the ablation takes place.
Because the velocity of the ablated plasma is
at most its sound velocity,
$v_{ab}\sim c_{s}\sim 10^7 \mbox{cm s}^{-1}$,
the ablated plasma expands only up to $\sim 10  \mu\mbox{m}$
during the laser irradiation $\sim 100 \mbox{ps}$.
This is much smaller than the scale of the system we consider
in this paper.
The laser energy deposited into thermal energy is
$E_{l}=526 \mbox{J}$ for all simulations.
Energy of ionization is not included,
therefore in a real experiment the incident laser energy,
of course, has to be significantly larger.

Four simulations have been done in this paper.
In cases 1 and 2, the ambient gas is very dilute 
($\rho _{a}=10^{-6} \mbox{g cm}^{-3}$).
In the first case, we neglect the radiation cooling,
while in the second case the radiation cooling is turned on.
These intend to simulate the experiments.
In the other two cases, 3 and 4, we let the plasma expansion in the
presence of an ambient gas.
Case 3 neglects the radiation cooling,
and case 4 is done with the radiation cooling term.
No corresponding experiments have been carried out as yet.
These models serve to clarify the effect of the ambient gas.

\begin{figure}
\rotatebox{270}{\resizebox{6.0cm}{!}{\plotone{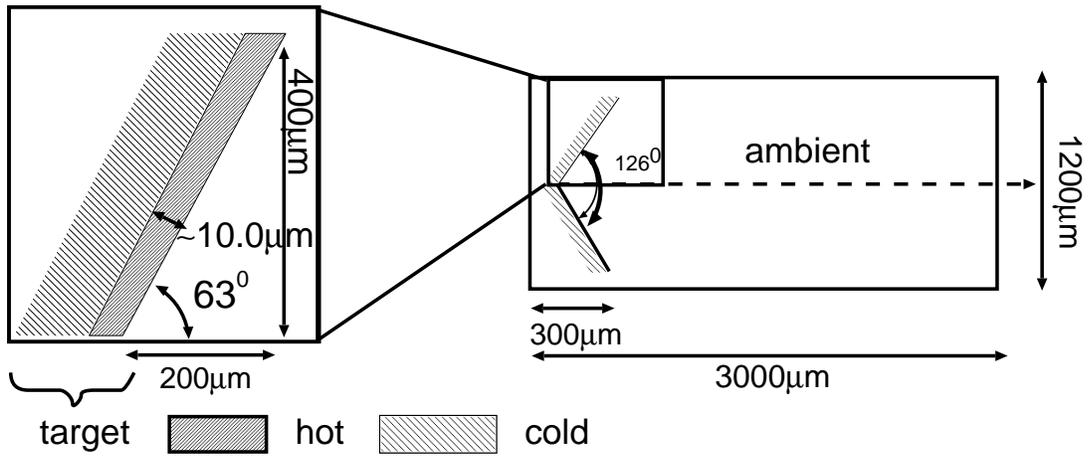}}}
\caption{Initial conditions. The cone target is set at
the left end of computational region
($1200 \mu\mbox{m}\times 3000 \mu\mbox{m}$).
Hot plasma is put at the surface. It's depth is $\sim 10 \mu\mbox{m}$.
\label{initialcd}}
\end{figure}

The temperature of the target surface,
where the laser energy is deposited,
increase to $\sim 400$ eV.
The angle of the cone is $126^{0}$,
which is similar to the experiments (see Fig.\ref{initialcd}).
The target matter is chosen to be gold (${}^{79}\mbox{Au}$) 
so that we can obtain large effect of the radiation cooling
due to a high-Z plasma.
The density of the target $\rho_{t}$
is initially uniform, $\rho _{t}=19.2 \mbox{g cm}^{-3}$,
and equals to the solid density of the gold.
Table \ref{tbl-1} summarizes the parameters of the initial conditions.

\section{Results}
\subsection{Vacuum case}
\begin{figure}
\resizebox{14.0cm}{!}{\plotone{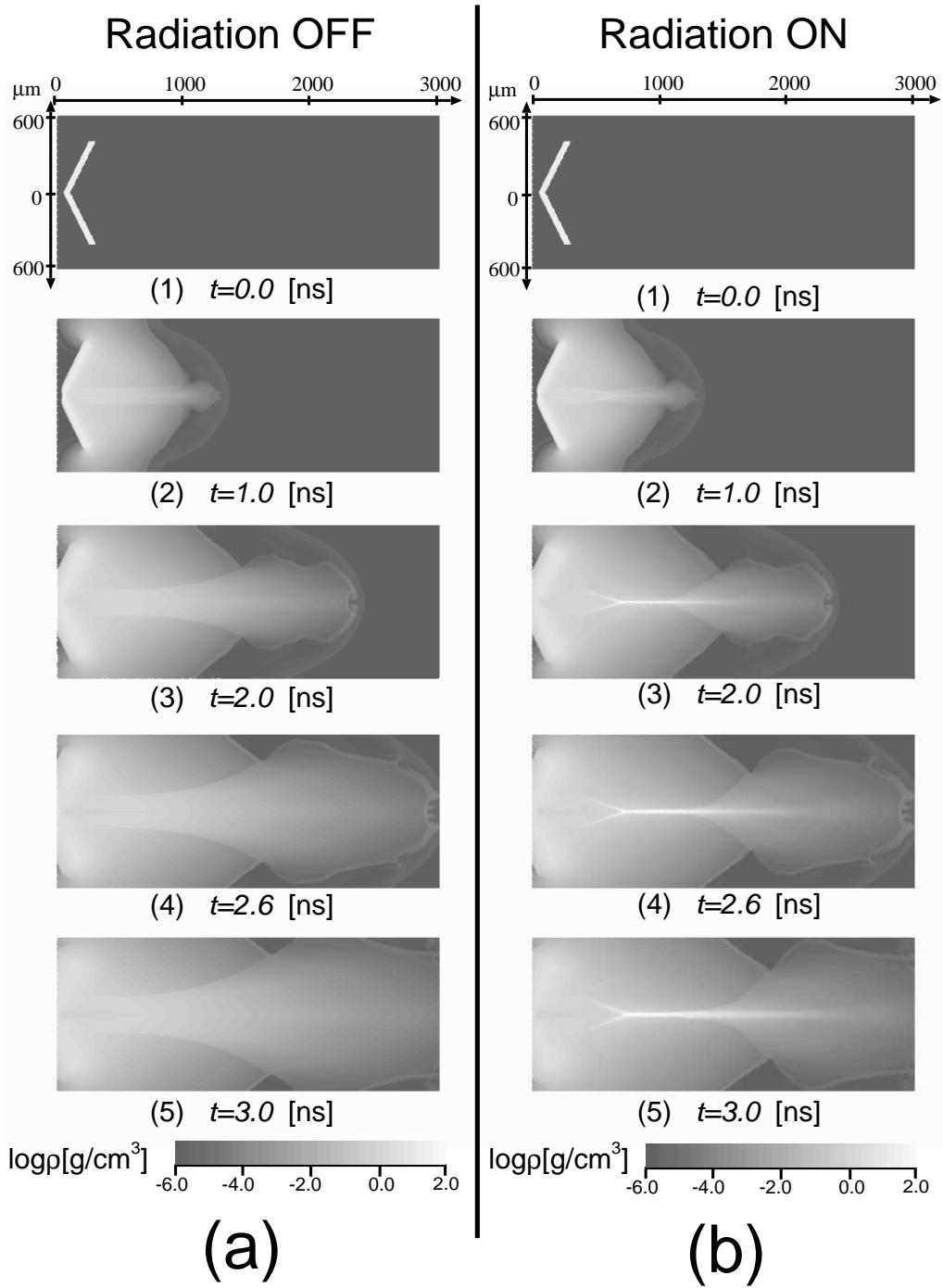}}
\caption{Density profiles in $r-z$ plane,
in the vacuum case;
(a)radiation off(case1),(b)radiation
 on(case2). $t=$0.0(1),1.0(2),2.0(3),2.6(4),3.0(5)$\mbox{ns}$
\label{rho6onoff}}
\end{figure}

First, we discuss cases 1 and 2 
which simulate plasma expansion into vacuum.
This condition is similar to the experiments of 
\citet{Far99,Shige00}.
For numerical reasons, we assumed in these cases as well a cold
($T_{a}= 0.026 \mbox{eV}$) and very low density
($\rho_{a}=10^{-6} \mbox{g cm}^{-3}$)
ambient gas.
Due to this very low density this ambient gas had no
influence on any of our results.
Figure \ref{rho6onoff} shows the density contours at different times with 
the radiation cooling off (Fig.\ref{rho6onoff}a)
and on (Fig.\ref{rho6onoff}b), respectively.

\begin{figure}
\rotatebox{270}{\resizebox{7.0cm}{!}{\plotone{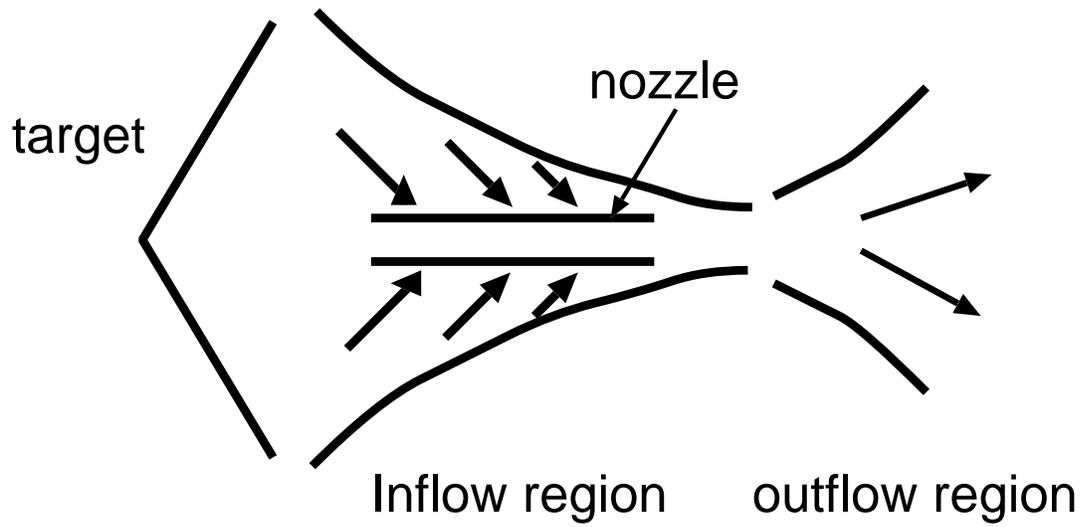}}}
\caption {Schematic flow structure. The flow is divided to two
 regions, inflow and outflow region.\label{region}}
\end{figure}

In the case of no radiation cooling (Fig.\ref{rho6onoff}a),
we can see two regions,
that is, the inflow region in which matter flows to the axis,
 and the outflow region.
This is schematically shown in Fig.\ref{region}.
Most of the laser-heated matter expands from the target, 
forming plasma with velocity of a few hundreds of kilometers per second,
farther away from the target it converges to the symmetry axis,
making a nozzle-like structure in the inflow region.
Then, the plasma in the nozzle spouts out 
from the tip of the nozzle, forming the outflow region.
Because the ambient matter is very dilute,
there is no pressure support to sustain this nozzle structure.
As a result, the outflow spreads out in all directions.
This flow structure is suitable for low Z targets,
because the effect of the radiation cooling is
indeed negligible.

\begin{figure}
\resizebox{14.0cm}{!}{\plotone{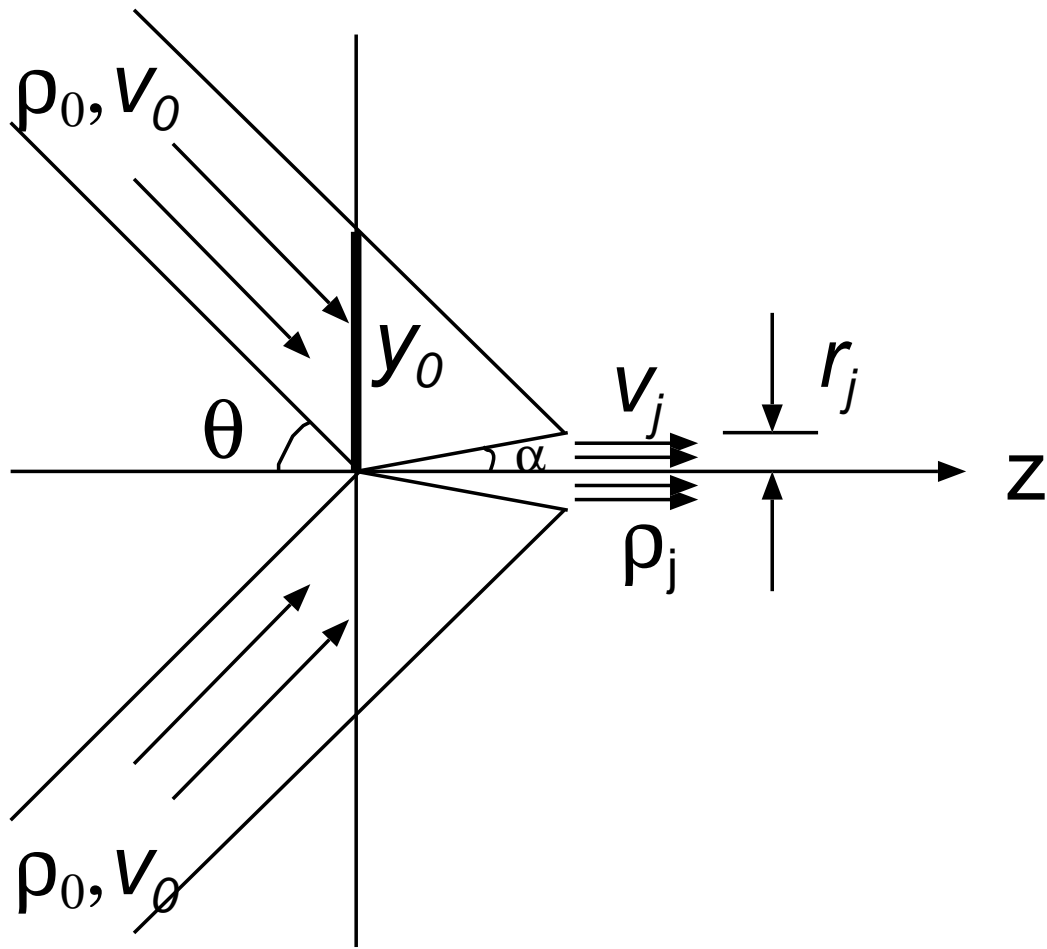}}
\caption{A schematic drawing of the conical flow to the central axis.
\label{conversion}}
\end{figure}

An analysis of the nozzle structure in the inflow region
is in order.
The converging flow from the cone target to the central axis is
very similar with the structure considered by \citet{Canto88}.
From their theoretical model we can estimate the beam radius $r_{j}$
and velocity $v_{j}$ from the converging flow angle $\theta$,
the reciprocal of the compression ratio
$\xi$ between the density of the nozzle to that of the inflow and
the velocity of the converging flow $v_{0}$
(see Fig.\ref{conversion}).
These are $\theta = 26.5^{0}$ and $\xi = 1/4$
in our case.
Then $r_{j}$ and $v_{j}$ are determined as
\begin{eqnarray}
r_{j}={\tan \alpha \over \tan \theta + \tan \alpha}y_{0},\\
v_{j}={\cos (\theta +\alpha)\over \cos \alpha}v_{0},
\end{eqnarray}
where $y_{0}$ is the width of the inflow
and $\alpha$ is the angle between the conical shock and z-axis
and given as
\begin{eqnarray}
\tan \alpha ={(1-\xi)+[(1-\xi)^2-4\xi\tan^2 \theta]^{1/2}
\over 2 \tan \theta}.
\end{eqnarray}
Applying this theory to our case with $y_{0}= 250 \mu \mbox{m}$
and $v_{0}\sim 3 \times 10^7 \mbox{km s}^{-1}$,
we obtain $r_{j}= 180\mu \mbox{m}$ and
$v_{j}\sim 1 \times 10^7 \mbox{km s}^{-1}$,
in good agreement with our numerical results.

For high Z targets, we have to consider radiation cooling effects. 
When radiation cooling is taken into account,
the flow structure is changed dramatically (Fig.\ref{rho6onoff}b).
The plasma in the  inflow region is collimated strongly
and shows a very thin nozzle at a later time 
(the radius of the structure is $\sim 40 \mu\mbox{m}$).
Although the collimation of the plasma is sustained in the inflow region
by converging inflow from the target,
the matter again spreads out in the outflow region,
resulting in almost the same structure as that for 
the case without radiation cooling.
These experiments are not suitable for the study
of the propagation of collimated jets.

\begin{figure}
\rotatebox{270}{\resizebox{6.5cm}{!}{\plotone{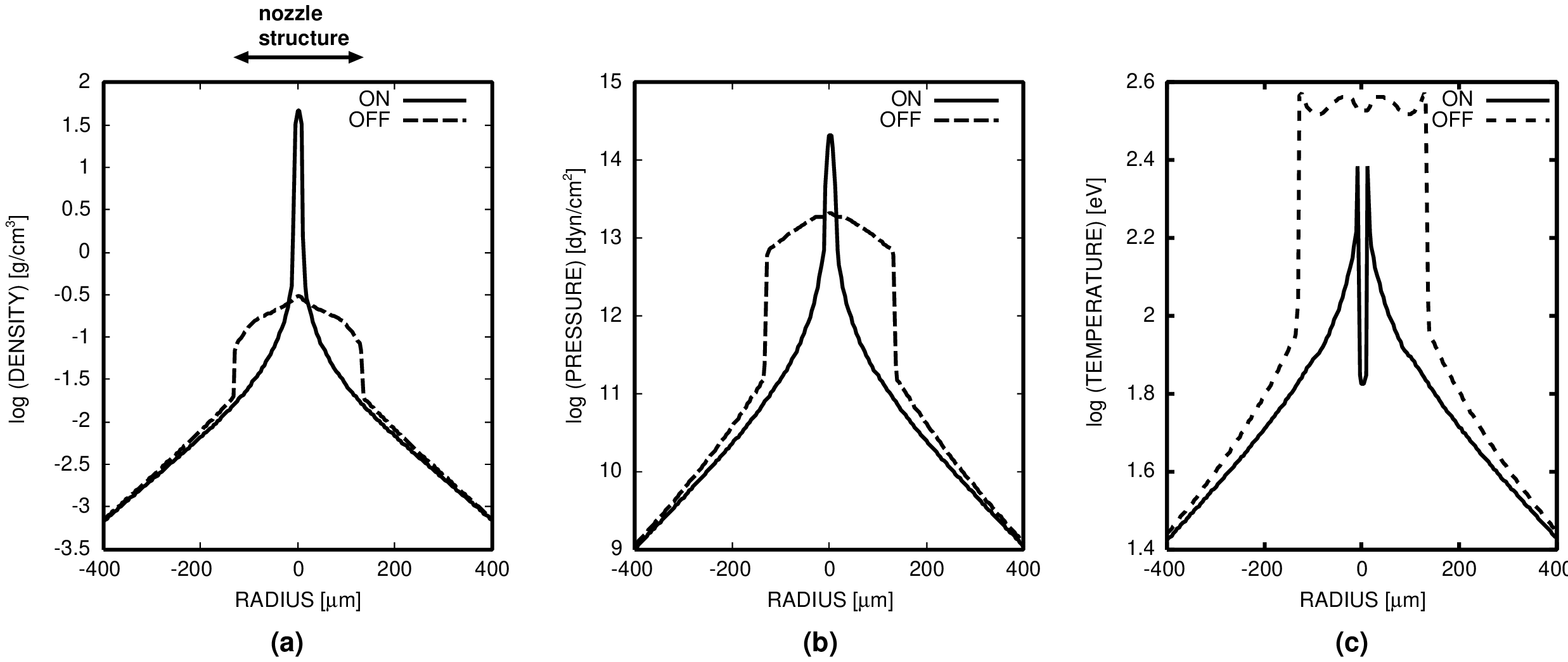}}}
\caption{Density, pressure and temperature profiles along r axis 
at z=1000$\mu$m, t=2.0ns. 
The radiation cooling term is OFF(dash line,case 1) and
ON(solid line,case 2).\label{rho6}}
\end{figure}

Figures \ref{rho6} a, b and c are the density, pressure 
and temperature profiles along $r $-axis, respectively,
at $z=1000 \mu\mbox{m}, t=2.0 \mbox{ns}$.
In the case with radiation cooling,
the plasma is cooled efficiently around the symmetry axis,
leading to the increase of the pressure gradient.
As the density increases, the plasma is further cooled,
and as a result a jet-like nozzle is formed,
which was actually observed in the experiments.
Altogether, our simulations support the experimental results
that the radiation cooling is a crucial ingredient
for the formation of this nozzle.

\subsection{Dense gas case}
\begin{figure}
\resizebox{14.0cm}{!}{\plotone{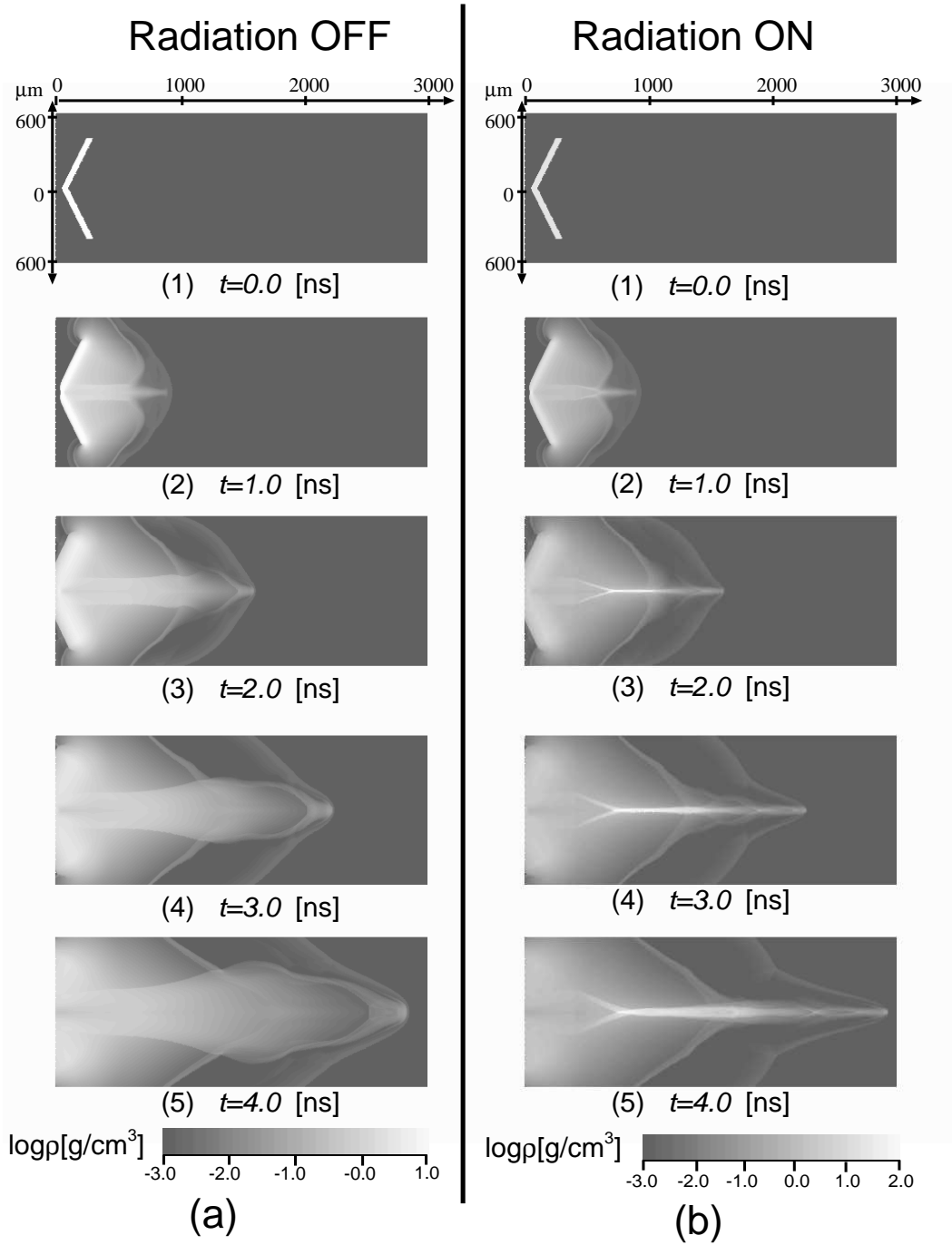}}
\caption{Density profiles in $r-z$ plane,
in the dense gas case;
(a)radiation off(case3),(b)radiation
 on(case4). $t=$0.0(1),1.0(2),2.0(3),3.0(4),4.0(5)$\mbox{ns}$
\label{rho3onoff}}
\end{figure}

In cases 3 and 4, the ambient matter density was much higher,
$\rho _{a}=10^{-3} \mbox{g cm}^{-3}$.
The time evolution of hydrodynamics leading to the formation
of the inflow and outflow regions is essentially unchanged.
Figure \ref{rho3onoff} shows the density contours at different times with 
the radiation cooling off (Fig.\ref{rho3onoff}a)
and on (Fig.\ref{rho3onoff}b), respectively.
The bow shock, which accelerates the ambient matter,
can be seen clearly.

In the case with radiation cooling,
the flow structure is dramatically changed as is obvious
in Fig.\ref{rho3onoff}b.
The effect of radiation cooling manifests itself in both
the inflow and the outflow regions.
In the inflow region, the nozzle structure becomes very thin just like
in the vacuum case in the later time.
The main difference appears in the outflow region.
Unlike all the other cases,
the width of the outflow region is very small.
Opposed to intuition,
this narrow structure is not a direct outcome of the thin nozzle
in the inflow region.
This is understood from the second panel of Fig.\ref{rho3onoff}b,
in which there is already a very thin outflow region
while there is no collimated structure in the inflow region.
When the matter enters the outflow region,
it tends to spread out as in vacuum case.
However, the shocked ambient matter slows the expansion 
in the direction perpendicular to the symmetry axis,
and the density and temperature of the flow remain high enough for
the radiation cooling to be effective.
As a result, a collimated plasma is produced in the outflow region.

\begin{figure}
\rotatebox{270}{\resizebox{6.5cm}{!}{\plotone{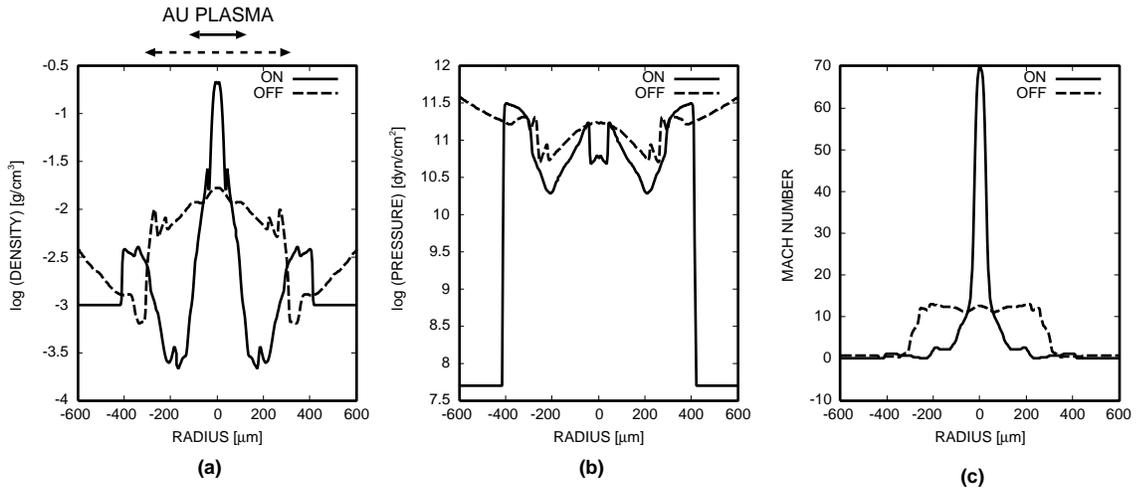}}} \\
\caption{Density, pressure and Mach number profiles along r axis 
at z=2000 $\mu$m, t=4.0 ns. 
The radiation cooling term is OFF(dash line,case 3) and
ON(solid line,case 4).\label{rho3}}
\end{figure}

This difference of structure in the outer region is
remarkable, as is further evident in Fig.\ref{rho3}, in which
the density, pressure 
and Mach number profiles along $r$-axis, respectively,
at $z=2000 \mu\mbox{m}, t=4.0 \mbox{ns}$ are shown.
The radius of jet is about 40$\mu\mbox m$,
and the density ratio is $\eta =\rho_{jet}/\rho_{a}\sim 100$,
namely, this is a so-called dense jet.
From Fig.\ref{rho3}b, we see that
the narrow structure in the outflow region is
sustained by the shocked ambient matter,
whose pressure is about 1000 times larger than
that of the unshocked ambient matter.
The jet is supersonic $M_{jet}\sim 70$.
The velocity of jet is $\sim 400 \mbox{km s}^{-1}$.
This is of the same order as the velocity of protostellar jets.
In order to study the propagation of astrophysical jets in the
laboratory,
we think it is necessary to do experiments with ambient gas.

\section{Discussion}
Firstly we consider the scaling law between a laboratory jet
and a protostellar jet.
The velocity of these two jets is almost
the same-a few hundred kilometers per second.
The ratio of the time scales is, however, widely different
$10^{2-4} \mbox{yr}$ for protostellar jets and
$10^{-9} \mbox{s}$ for laboratory jets,
a scale up of a factor of $10^{18-20}$.
For the experiment to simulate efficiently
the astrophysical jet,
the ratio of the lengths of jets has to be $\sim 10^{18-20}$.
The typical ratio of the radii of the jets ($r_{prot}/r_{lab}$) is
$10^{15} \mbox{cm}/10^{-3} \mbox{cm}=10^{18}$.
Thus, the aspect ratios of the jets are close to each other.
The temperature is a few eV for both jets,
which is consistent with the scaling of velocity mentioned above.
Thus, a good scaling law holds between the protstellar jet and
the laboratory jet considered in this paper.

The relevant parameters which characterize the jet are
the density ratio $\eta $, the Mach number and velocity of jets
$M_{jet},v_{jet}$, the pressure ratio $K$ and the cooling parameter $\chi$.
As the magnetic fields were not included our code,
no comparison can be made with magnetohydrodynamic simulations
\citep{Todo93,Cerqueira97,Frank98,Gardiner00,Frank00}.
As is obvious from the results obtained
both for the vacuum and the ambient gas cases,
the radiation cooling is the most crucial ingredient
in the collimation of the jets.

The cooling parameter is defined from the cooling term
which has already been employed in the hydrodynanmic simulations,
$|J_{rad}|$ in  Eq.(\ref{cooling}).
It is the ratio
\begin{eqnarray}
\chi ={\mbox{cooling length}\over
 \mbox{jet radius}}\sim 
{v_{jet}\tau_{rad}\over r_{jet}},
\end{eqnarray}
where $v_{jet}$ and $r_{jet}$ are the velocity and radius
of jet, respectively,
and $\tau_{rad}$ is a characteristic radiation cooling time.
$\tau _{rad}$ is defined from the ratio of thermal energy density to
emission power as
\begin{eqnarray}
\tau_{rad}={e_{thr}\over |J_{rad}|}.
\end{eqnarray}
The smaller $\chi$ is, the more effective the cooling is.
The cooling term adopted in Eq.(\ref{cooling}) is different
from those used in other papers \citep{Blondin90,Stone93}.
We assumed in this paper that the jet matter is Au,
rather than hydrogen in most other papers.
Our choice was influenced by the need
to enhance the radiation cooling so that
the hydrodynamics is affected within a few nanoseconds.
In fact, the average ionic charge state $Z^{*}$ is
about $20-40$ for $T=50-100 \mbox{eV}$.
Although the detailed dependence on $T$ is different between our
cooling term and others, it is emphasized that the
dynamics should be similar as long as the cooling parameter
$\chi$ has a similar value during the greater part of the time evolution.

In case 4 (with dense ambient gas and radiation cooling),
the typical radiation cooling time is 
$\tau_{rad }\sim 10 \mbox{ns}$ at $t=1.0 \mbox{ns}$,
and $\tau_{rad }\sim 100 \mbox{ns}$ for $t=4.0 \mbox{ns}$.
These were evaluated at the points with the largest emission power.
Accordingly, the radiation cooling parameters for each time
are $\chi \sim 10$ and $\chi \sim 100$, respectively.
Thus, the radiation cooling is effective for the earlier phase,
while the jet is cool enough, while the radiation cooling is ineffective
for the later phase.
However, it is noted that we underestimate
the radiation cooling,
since we neglected the line emissions
which could be substantial in our cases,
and take into account only the bremsstrahlung.
This was done mainly to keep numerical simplicity.

The aim of our calculations and future experiments is
to understand better the physics of protostellar jets.
Although the observational values of physical quantities
are rather uncertain,
they are typically,
$n_{e}\sim 10^3 \mbox{cm}^{-3}$, $T_{j}\sim 10^4 \mbox{K}$,
$v_{j}\sim 100 \mbox{km s}^{-1}$ and $r_{j}=10^{15} \mbox{cm}$
\citep{Bally01}.
These values roughly correspond to
$\eta \approx 1- 10^3, M_{jet} \approx 1$
with the main uncertainty coming from the lack of data
on the density and temperature of the ambient matter.
From these parameters,
it is estimated that the total emission power is 
$|J_{rad}|\sim 10^{-18} \mbox{erg cm}^{-3}\mbox{s}^{-1}$
for collisional by excited hydrogen atoms \citep{Dar72}.
The emission power due to the bremsstrahlung is
$|J_{rad}|\sim 10^{-19}\mbox{erg cm}^{-3}\mbox{s}^{-1}$.
Thus the bremsstrahlung is substantially weaker,
and in general the bound-bound emission is dominant in such plasmas.
The emission power $|J_{rad}|\sim 10^{-18}\mbox{erg cm}^{-3}\mbox{s}^{-1}$
gives a cooling time of $\tau_{rad}\sim 10^9 \mbox{s}$,
which then leads to a cooling parameter of $\chi \sim 10$.
It is evident from these values that the experimental values
used in this paper
are appropriate for studying astrophysical jets.
As mentioned in the introduction,
\citet{Blondin90,Stone93}
studied the properties of propagating jets
by varying $\eta$ between 1 and 10.
Their parameters correspond to $\chi$ between 0.1 and 10.
We think that our results have overlapping region
to the initial conditions of those papers.
This means that the experiments proposed here
might give  some new insights into the
astrophysical jets studied theoretically in their papers.

\section{Conclusion}
In this paper results of
four hydrodynamic simulations have been shown,
in which jet were generated
from 'cone' target irradiated by intense laser.
In all the simulations,
we have found a common feature,
that is, the existence of an inflow region and an outflow region.
Depending on whether the radiation cooling is turned on or off,
and whether there is some ambient gas or not,
the flow structure inside these regions is very different from each other.

When the plasma flows into a very low density ambient gas,
the collimation of the plasma jet occurs only due to radiation cooling.
With radiation cooling,
the collimated plasma shows up in the inflow region.
However, since the structure is
sustained by converging inflow from the target,
it disappears in the outflow region,
in which there is no pressure support to maintain it.
This feature is very similar to that found in the experiments.
We have also found that the nozzle structure with no radiation cooling
is well described with the analytical model by \citet{Canto88}

In the case with ambient gas,
a bow shock appears very clearly.
The main difference between the vacuum and dense gas cases is
that in the latter case the collimated flow
appears not only in the inflow region
but also in the outflow region.
Particularly, the collimated plasma in the outflow region
is sustained by the pressure of the shocked ambient matter.

We have shown the possibility to carry out a laboratory experiment,
using high intensity laser-plasma interaction,
to generate collimated plasma jets propagating into an ambient gas.
The parameters of such plasma jet are similar to those of
protostellar jets.
If such an experiment is performed,
we can get important information about the physics of
astrophysical jets.

External magnetic fields were not included in our calculations.
Experiments with magnetic fields have already proposed
and will be carried out in the future.
We intend to adjust our simulations
to the needs of future experiments.

\

This work was carried out on NEC SX4, Cybermedia Center Osaka Univ. and
HITACHI SR8000, Institute of Laser Engineering, Osaka Univ.

One of authors (A. Mizuta) would like to thank K. Sawada, H. Nagatomo 
and N. Ohnishi for good suggestions for numerical methods.

\clearpage

\begin{table}
\begin{center}
\caption{Detailed parameters for initial condition. $\rho$ and $T$ mean
density and temperature. Index 't' and 'a'
stand for target and ambient matter\label{tbl-1}}
 \begin{tabular}{c|cccc}
\tableline\tableline
 &  case1 & case2 & case3 & case4\\ \hline
target &\multicolumn{4}{c}{Au}\\
$\rho_{t} \mbox{g cm}^{-3} $ &\multicolumn{4}{c}{$19.2$}\\
$T_{t} \mbox{eV}$(hot/cold) &\multicolumn{4}{c}{$676/1.77$}\\
$\rho_{a} \mbox{g cm}^{-3}$ &\multicolumn{2}{c}{$10^{-6}$}
&\multicolumn{2}{c}{$10^{-3}$}\\
$T_{a}[\mbox{eV}]$ &\multicolumn{4}{c}{$0.026$}\\
cooling term & OFF & ON & OFF & ON\\ \tableline
\end{tabular}
\end{center}
\label{initable}
\end{table}

\clearpage

\clearpage

\clearpage

\clearpage

\clearpage

\clearpage

\clearpage

\clearpage

\end{document}